\documentclass{article}
\usepackage{comment}
     \usepackage[final]{neurips_2020}

\usepackage[utf8]{inputenc} % allow utf-8 input
\usepackage[T1]{fontenc}    % use 8-bit T1 fonts
\usepackage{hyperref}       % hyperlinks
\usepackage{comment}       % hyperlinks
\usepackage{bibunits}
\usepackage{graphicx}
\usepackage{setspace}
\usepackage{amsmath}
\usepackage{url}            % simple URL typesetting
\usepackage{booktabs}       % professional-quality tables
\usepackage{amsfonts}       % blackboard math symbols
\usepackage{nicefrac}       % compact symbols for 1/2, etc.
\usepackage{microtype}      % microtypography
\title{The Managerial Effects \\of Algorithmic Fairness Activism}

\author{%
  Bo Cowgill\\
  Columbia Business School, Columbia University\\
  \texttt{bo.cowgill@gsb.columbia.edu} 
    \And
  Fabrizio Dell'Acqua\\
  Columbia Business School, Columbia University\\
  \texttt{fdellacqua21@gsb.columbia.edu} 
    \And 
  Sandra Matz\\
  Columbia Business School, Columbia University\\
  \texttt{sm4409@gsb.columbia.edu} 
}

\begin{document}

\maketitle

\begin{abstract}
  How do ethical arguments affect AI adoption in business? We randomly expose business decision-makers to arguments used in AI fairness activism. Arguments emphasizing the inescapability of algorithmic bias lead managers to abandon AI for manual review by humans and report greater expectations about lawsuits and negative PR. These effects persist even when AI lowers gender and racial disparities and when engineering investments to address AI fairness are feasible. Emphasis on status quo comparisons yields opposite effects. We also measure the effects of ``scientific veneer'' in AI ethics arguments. Scientific veneer changes managerial behavior but does not asymmetrically benefit favorable (versus critical) AI activism.
\end{abstract}

  \defaultbibliography{algorithmic-fairness.bib}
  \defaultbibliographystyle{aer}
\begin{bibunit}
\emph{Our submission is a 4-page summary to comply with the workshop page limits. Our full paper can be viewed at \href{https://www.aeaweb.org/articles?id=10.1257/pandp.20201035}{https://www.aeaweb.org/articles?id=10.1257/pandp.20201035}.}

\section{Introduction}
In the last five years, the ethical and distributional consequences of algorithmic decision-making have become a rising concern. Interest in ``algorithmic bias,'' measured by Google Trends, has increased more than ten-fold during this time. Bias in algorithms is now the topic of major regulatory efforts, new lines of academic research, and even political debates. How has this surge of attention to fairness considerations affected the adoption of AI? 

In this paper, we assess this question by surveying business decision-makers.\footnote{Related research surveys ML practitioners and decision-makers, see for example, \cite{rakova20,holstein19}. While these conduct semi-structured interviews, our studies experimentally manipulate the arguments in favor or against AI that the practitioners read.}  Our surveys include experimental manipulations of arguments about algorithmic fairness. We randomly expose business decision-makers to arguments used in AI fairness activism. We then measure the effects on AI adoption recommendations and supporting beliefs. In particular, we assess two broad sets of interventions. 

We first examine the effect of arguments about the origin of algorithmic bias and its relationship to the status quo. In our experiment, subjects are randomly assigned to read an op-ed. One op-ed emphasizes that algorithmic bias is unavoidable. We refer to this as our ``AI fatalism'' condition. The second op-ed claims that while algorithmic bias exists, algorithmic decision-making might be preferable to human-based alternatives which could possibly contain higher levels of bias. We refer to this as the ``counterfactual'' condition. 

In our second study, we measure the effect of adding ``scientific veneer'' to arguments about AI ethics using a 2$\times$2 design. Subjects are randomly assigned to read arguments highlighting either the favorable or unfavorable ethical assessments of AI systems. These arguments are presented either with scientific veneer or not (randomly assigned). We constructed scientific veneer treatments to attach symbols of scientific authority to arguments, but without introducing new substantive content to the claims. Our experimental materials can be seen in Appendix \ref{MaterialsVeneer}. 

Both studies measure adoption of AI in settings motivated by \cite{cowgill2019bias} (hiring) \cite{bartlett2019consumer} (lending). We assess not only adoption decisions, but also a series of potential mechanisms behind these decisions (specifically beliefs about the probability of fairness PR or legal allegations, and the magnitude of these problems conditional on arising). We also measure beliefs about the ``fixability'' of algorithmic unfairness problems in response to focused engineering efforts. Finally, we examine how information about status quo processes affects choices about AI.

\noindent {\bf Main Findings}: In our first study, we find that our ``fatalism''' op-ed discouraged AI adoption, while the ``counterfactual'' encourages it (compared to a baseline of no op-ed). This effect seems to be driven by differences in the perceived likelihood of being caught in a lawsuit or PR problem. The perceived magnitude of such a problem -- in addition to the probability of one existing at all -- is also substantially higher under the ``fatalism'' condition (than under the``counterfactual'' condition) but the effect is larger on the probability of detection. 

Regarding the fixability of the algorithmic biases, our business decision-makers believe fairness problems are more correctable under our ``fatalism'' conditions. Lastly, we find that ``fatalism'' arguments discourage counterfactual thinking in our subjects. When given the option, subjects are less likely to investigate comparisons between adopting AI and the status quo, to make decisions (mostly rejecting AI) without examining alternatives. 

In our second study, we find some evidence (albeit limited) that ``scientific veneer'' enhances arguments promoting the ethical properties of AI. However, we find evidence that scientific veneer amplifies arguments \emph{against} the ethical properties of the AI. This is particularly true with regards to beliefs about the fixability of AI ethics issues. 

Math-based arguments were found to have a heterogeneous effect on counterfactual thinking. Adding mathematical veneer to positive arguments caused subjects to be more likely to investigate the status quo. However, adding math veneer to negative arguments \emph{lowered} levels of counterfactual thinking, comparing and conditioning on the status quo. 

This article contributes to a broader literature on adoption and effects of algorithmic decision-making in economics \citep{cowgill2019economics} and psychology \citep{dietvorst2015algorithm}, as well as on ethical AI considerations by firms \citep{rakova20,holstein19,madaio20}   Our results have mixed implications for strategies for making business decision-makers mitigate bias in AI. The remainder of this paper proceeds as follows. In Section \ref{ExperimentalDesign} we describe the design of our two studies. In Section \ref{Results} we review results. In our conclusion (\ref{Conclusions}), we summarize and discuss the implications of our findings. 

\section{Experimental Design}\label{ExperimentalDesign}

Our survey was conducted on January 8, 2020, using Prolific Academic.\footnote{Evaluated in \cite{peer2017beyond}} We focused our study on subjects likely to make AI adoption decisions for organizations. We, therefore, limited participation to U.S. adults with management experience. Table \ref{tab:DescStats} reports descriptive statistics for our sample.

Both our studies took the following format. Each subject reads two short business case studies. Each case puts the subject in the role of a manager making an AI adoption decision. The first case is about using AI for hiring at a technology company, and the second is about using AI for lending decisions by financial institutions. Each case concludes by disclosing an imbalance in the algorithm's behavior along gender (Study 1) or racial (Study 2) lines. The level of imbalances are taken from \cite{cowgill2019bias} (hiring) \citep{bartlett2019consumer} (lending). The text of both cases is in Appendix \ref{CaseStudyTexts}; we describe key features below. 

After each case, subjects are exposed to information about the level of imbalances for the status quo process; i.e., what happens if the AI is rejected. In both of these papers, imbalances appear in both algorithmic and status-quo processes, but are \emph{lower} in the algorithmic condition. A random third of subjects are shown this status quo data. For a second third, the status quo information is available only if the subjects are willing to click to view it (rather than proceeding to questions without viewing it). For the final random third, the status quo is not mentioned at all. 

Subjects then proceed to answer questions about the case. Each subject answers five adoption-related questions. The first asks how positively subjects consider the impact of this technology. The second and third ask for a recommendation to adopt or maintain the status quo. The fourth asks how likely lawsuits or PR problems are if the algorithm is adopted. The fifth asks about the magnitude of lawsuits or PR problems, conditional on them existing. After these five questions, the subject is told to imagine that the company in the case has dedicated six months of additional engineering effort to address the imbalances at the end of the case. Subjects are then asked to re-answer the five questions, incorporating beliefs about how the algorithm would change after six months of dedicated programming effort. 

All subjects go through the above procedure twice, once for hiring and once for lending. For all subjects, the second case study (in randomized order) features an additional randomized treatment: a \emph{persuasive argument} about algorithmic fairness. We use this treatment to study how various types of arguments change subjects' adoption decisions (and their beliefs surrounding these decisions).  We describe these persuasive interventions below. Both studies were allocated $\approx$500 subjects. 

\noindent {\bf Study 1: ``Counterfactual and Fatalistic Perspectives''}: In our first study, subjects are randomly assigned into three conditions to read op-eds in between the first and second business case. A recent media article\footnote{\url{https://www.axios.com/ai-bias-c7bf3397-a870-4152-9395-83b6bf1e6a67.html}, accessed Jan 9, 2020.} described a split in perspectives around AI ethics (``A tug-of-war over biased AI''). We chose an op-ed reflective of each perspective, and randomly assigned one third of subjects to read it between the two cases. The final third read no op-ed. 

The first op-ed conveys the perspective that AI cannot be fully cleansed of bias and thus should be avoided altogether. We refer to this as the ``fatalistic'' perspective on AI bias. In this condition, subjects read ``A.I. Bias Isn’t the Problem. Our Society Is,'' an essay published in Fortune magazine.\footnote{\url{https://fortune.com/2019/04/14/ai-artificial-intelligence-bias/}, accessed Jan 9, 2020.} The second op-ed reflected the view that AI may exhibit biases, but can still be useful if the biases are reduced below the levels of counterfactual human decision-making. We refer to this as the ``counterfactual'' perspective as it often emphasizes comparisons against the status quo. In this condition, subjects read ``Want Less-Biased Decisions? Use Algorithms,'' an essay published in Harvard Business Review.\footnote{\url{https://hbr.org/2018/07/want-less-biased-decisions-use-algorithms}, accessed Jan 9, 2020.} 

\noindent {\bf Study 2: ``Scientific Veneer''}: In our second study, we examine a related set of hypotheses about persuasion in AI. This perspective appears in many critical outlets, but was succinctly articulated by U.S. Senator Kamala Harris, who wrote that AI could ``hide bias behind a scientific veneer of objectivity.''

These critics alleged that AI advocates benefit from the use of statistics and engineering in their application and persuasion. They also allege that the biographies of AI engineers give them scientific authority, even though ``Algorithms are opinions embedded in code,'' according to one commentator.\footnote{See \emph{Weapons of Math Destruction} by Cathy O'Neil.} However, the persuasion of AI critics also exhibits mathematical arguments and credentials. Leading critics include computer science professors, statisticians, engineering executives, and math PhDs. These features may similarly lend a veneer of objectivity to critical arguments about AI ethics. 
% COMMENTED OUT Some psychological theories \citep{} suggest that they might, in fact, have a disproportionately persuasive effect. 

We study the effects of scientific veneer on both negative and positive arguments regarding the ethics of AI. In Study 2, subjects read a report from an expert between the two cases. The report appears after the second business case before the subject examines the status quo (if available) and before questions about adoption.  The content of the report is randomized to present either negative or positive arguments about the ethics or bias of the AI application. In addition, the report was randomized to include scientific veneer or not. The text of the ``veneer'' report is identical to the non-veneer one, but also repeats the verbal statements using mathematical notation.\footnote{Our ``veneer'' interventions were carefully constructed for redundancy; i.e., \emph{not} to include any new arguments. Examples appear in Appendix \ref{MaterialsVeneer}.}

Our veneer treatments also differ in the expert's identity and the typesetting of the report. Both experts were U.C. Berkeley PhDs employed to evaluate algorithms. The ``veneer'' expert was a physics Ph.D., and the ``non-veneer'' expert was a sociology Ph.D. The physicist's report was created in \LaTeX (a typesetting program popular in mathematics, natural sciences, and engineering). The sociologist's report was prepared in Microsoft Word using Arial font. 

\section{Results}\label{Results} We summarize results verbally below, but Appendix \ref{Specifications} provides the specifications in our results tables (Appendix \ref{TablesSection}) and some guidance on how to interpret the coefficients. Because each subject answered all questions twice, we employ subject fixed effects and panel methods. Note that positive coefficients refer to favorable evaluations of the algorithm, except for the ``lawsuit'' and ``damages'' column. All outcome variables are standardized. 

%Before introducing our results about AI fairness activism, we briefly summarize several consistent patterns documented in Tables \ref{tab:SqDemogsStudy1} and \ref{tab:SqDemogsStudy2}. Subjects forced to view the status quo (or who have the option) assess the algorithm more favorably along a variety of dimensions. Managers who make AI-related decisions on their jobs, and who feel educationally prepared for these decisions are also more favorable. Women, African-Americans, and political liberal managers are less favorable towards the algorithms. These results are much weaker when subjects are shown the status quo. In addition, we find strong results in all specifications that six months of focused engineering efforts will be effective. 

We report our results from Study 1 (``Counterfactual'' and ``Fatalism'' op-eds) in Table \ref{tab:EffectsOfRhetoricStudy1}. Reading the counterfactual op-ed gives a more positive perception of the algorithm and encourages the adoption of AI. It increases the perceived positive impact of the technology by 0.24 standard deviations and increases recommendations for adoption by 0.19 standard deviations. 

By contrast, reading the fatalistic op-ed leads to a more negative perception of the algorithm (-0.16 standard deviations), and makes adoption less likely, in particular when measured on a scale (-0.28 standard deviations). A relevant mechanism for these results seems to be connected to PR and potential lawsuits. Business decision-makers who read the counterfactual op-ed believe fairness lawsuits and PR problems are less probable (-0.12 standard deviations). At the same time, they also think that, if a problem was assessed, it would be less damaging (-0.14). Instead, subjects who read the fatalistic op-ed consider lawsuits more likely and at the same time more damaging, respectively by 0.19 and 0.061 of a standard deviation.

When we examine fixability outcomes, we observe that the counterfactual op-ed makes business decision-makers think that fairness problems have less room to be fixed compared to the no op-ed case. This may partly be the result of having set initial expectations relatively high in the counterfactual condition. The effects of the fatalism op-ed about fixability are mixed; in some specifications, they are positive but not statistically significant. Even after accounting for potential fixability effects, the net effect of the fatalism op-ed on recommendation is lower than if the subject read no op-ed at all. 

As mentioned above, subjects exposed to the status quo rated AI as more positive, and consider lawsuits less likely and less damaging. In Panel B of Table \ref{tab:EffectsOfRhetoricStudy1}, we focus on subjects who are given the \emph{option} to evaluate the status quo in comparison with the algorithm in the case study for a small price (an additional click). Using an IV specification, we observe that decision-makers who read the fatalistic op-ed were less likely to condition on them. Some of our specifications suggest that subjects who read the fatalism op-ed are less interested in finding out about the status quo at all. By contrast, subjects in the counterfactual condition appeared slightly more likely to explore the status quo and to condition choices on them. 

In Table \ref{tab:EffectsOfRhetoricStudy2} we present the results of the second study of ``scientific veneer.'' We find strong evidence that scientific veneer affects the decisions to adopt AI. After reading the same opinion with the addition of math symbols and jargon, decision-makers recommended algorithm’s adoption less by 0.21 standard deviations if this opinion is against AI. Instead, they recommended the AI more by 0.27 standard deviations when the opinion is in favor of AI. 

Across specifications and outcome variables, the magnitude of the effects of scientific veneer is approximately equal, irrespective of whether the veneer is applied towards favorable or unfavorable opinions in algorithms. Panel B of table \ref{tab:EffectsOfRhetoricStudy2} reports the results on counterfactual thinking. When decision-makers are exposed to positive arguments with scientific veneer, they are more likely to be interested in the status quo. They show less interest in the status quo when shown negative arguments with scientific veneer. 

\section{Conclusion}\label{Conclusions}

A key question for activists and policymakers is: What is the goal of AI fairness activism? If the goal is for organizations to abandon AI -- which some critics advocate -- the ``fatalistic'' perspective is more effective. We show that the abandonment effect persists even in settings like \citep{bartlett2019consumer} and \cite{cowgill2019bias} where algorithms reduce racial- and gender- disparities. The abandonment effect is also robust to the possibility of dedicated engineering improvements. 

If the goal is to encourage adoption when AI reduces disparities, and improving algorithmic bias when it doesn't -- we find that counterfactual arguments are more effective. This may be the best activism strategy in settings where human biases exist, but machine biases could feasibly reduce them. Additionally, our results can inform firms about how to incorporate AI ethics more effectively in their operations.\footnote{\cite{cowgillfabizio2020} shows simple reminders about algorithmic bias are among the most effective ways to encourage engineers to improve fairness in a hiring algorithm for a firm. Other common methods within companies include AI ethics checklists, discussed in \cite{madaio20}.}.

Our paper highlights the perspective of firms, for whom AI adoption features both benefits as well as costs. Our study finds that AI abandonment is correlated with fear of lawsuits and negative PR. In our setting, this results in decisions made by human review, which other researchers have shown to exhibit bias. Because human judgments are harder to audit, this shift may reduce lawsuit risk rather than reducing disparities.

\putbib
\end{bibunit}
\begin{bibunit}
\appendix
\clearpage
\section{Appendix: Tables and Figures}\label{TablesSection}

\begin{table}[h!]
 \caption{\textbf{Descriptive Statistics: Subjects}}
	 \label{tab:DescStats}
	% \centering
	\scriptsize
\centering
	\begin{tabular}{l*{14}{c}}
                    &       Op-Ed&      Veneer\\
\hline
Male                &        0.49&        0.52\\
Female              &        0.49&        0.45\\
Other Gender        &        0.02&        0.02\\
Latinx              &        0.10&        0.07\\
White               &        0.81&        0.85\\
Black               &        0.07&        0.05\\
Asian               &        0.05&        0.04\\
Other Ethnicity     &        0.06&        0.06\\
AI Decisions        &        0.49&        0.40\\
Prepared by Educ.   &        0.75&        0.71\\
Knows ML            &        0.46&        0.45\\
\end{tabular}

	 \flushleft \begin{footnotesize}
	 \begin{singlespace}

\textbf{Notes}:  All reported variables are binary. Subjects self identified their gender as ``Male'', ``Female'' or ``Other''. They selected their ethnic background, and whether they are Latinx. ``AI Decisions'' captures subjects who report working (or potentially working) in roles where they make decisions like those in our surveys. ``Prepared by education'' indicates whether subjects feel their education has prepared them well enough for this type of decision. ``Knows ML'' takes the value 1 if subjects know ``a great deal'', ``a lot'', or ``a moderate amount'' about machine learning and predictive modeling, and 0 otherwise. Subjects were recruited from Prolific Academic, which is evaluated in \cite{peer2017beyond}.  
	
	%\smallskip * significant at 10\%; ** significant at 5\%; *** significant at 1\%. 
	\end{singlespace}
	\end{footnotesize}
	\end{table}%

\begin{table}[h!]
\caption{\textbf{Status Quo Conditions and Demographics: Study 1 (``Counterfactual'' and ``Fatalism'' Activism)}}
	 \label{tab:SqDemogsStudy1}
	% \centering
	\scriptsize

\begin{tabular}{l*{14}{c}}
                    &\multicolumn{1}{c}{Positive}&\multicolumn{1}{c}{Rec (Scale)}&\multicolumn{1}{c}{Rec (Y/N)}&\multicolumn{1}{c}{Lawsuit}&\multicolumn{1}{c}{Damage Size}\\
\hline
Algorithm Fix       &         .47***&         .45***&     1.3e-14   &         -.7***&        -.34***\\
                    &      (.057)   &      (.051)   &         (.)   &      (.061)   &      (.044)   \\
Status Quo Shown    &         .52***&         .65***&         .57***&        -.43***&        -.44***\\
                    &        (.1)   &        (.1)   &      (.094)   &      (.095)   &      (.099)   \\
Status Quo Shown (only if clicked)&         .46***&         .63***&         .54***&        -.38***&        -.42***\\
                    &       (.11)   &      (.099)   &      (.096)   &      (.092)   &      (.098)   \\
Fix $\times$ Status Quo Shown&        -.17** &        -.26***&    -2.2e-14   &         .19** &         .12*  \\
                    &      (.072)   &      (.063)   &         (.)   &      (.084)   &      (.061)   \\
Fix $\times$  Status Quo Shown (only if clicked)&        -.19** &         -.3***&    -1.9e-14   &          .2** &        .091   \\
                    &      (.076)   &      (.062)   &         (.)   &      (.082)   &      (.061)   \\
Female              &        -.16** &        -.11   &        -.15** &         .11   &         .29***\\
                    &       (.07)   &      (.072)   &      (.071)   &      (.066)   &      (.071)   \\
Black               &        -.25*  &        -.22*  &       -.048   &         .18   &         .22*  \\
                    &       (.14)   &       (.13)   &       (.13)   &       (.11)   &       (.13)   \\
Asian               &        -.17   &       -.027   &       -.084   &       -.026   &        -.16   \\
                    &       (.13)   &       (.13)   &       (.16)   &       (.14)   &       (.15)   \\
Other Ethnicity     &        -.33***&        -.26*  &        -.21   &        .083   &       -.086   \\
                    &       (.12)   &       (.14)   &       (.16)   &       (.14)   &       (.15)   \\
Political Conservatism (Standardized)&        .094** &        .082** &        .026   &       -.067** &       -.079** \\
                    &      (.038)   &      (.037)   &      (.036)   &      (.034)   &      (.038)   \\
AI Decisions        &         .12*  &         .12*  &        .078   &        .025   &        .019   \\
                    &      (.069)   &      (.072)   &      (.075)   &      (.067)   &      (.074)   \\
Prepared by Educ.   &         .13   &         .14*  &        .099   &        -.12*  &        -.12   \\
                    &       (.08)   &      (.083)   &      (.082)   &      (.075)   &       (.08)   \\
Knows ML            &       .0055   &       .0028   &       -.056   &        .062   &       -.097   \\
                    &       (.07)   &      (.074)   &      (.074)   &      (.068)   &      (.076)   \\
\hline
Observations        &       1,992   &       1,992   &       1,992   &       1,992   &       1,992   \\
\(R^{2}\)           &          .1   &         .12   &         .12   &         .12   &        .097   \\
\hline $p$-value: Status Quo (SQ) Shown + Fix $\times$ SQ Shown=0&       .0005   &      .00018   &     2.4e-09   &        .024   &       .0023   \\
$p$-value: SQ Optional + Fix $\times$ SQ Opt=0&       .0075   &       .0018   &     3.0e-08   &        .097   &       .0017   \\
\hline $p$-value: SQ Shown = SQ Opt&         .48   &         .85   &         .74   &         .57   &         .85   \\
$p$-value: Fix $\times$ SQ Shown = Fix $\times$ SQ Opt&         .75   &         .37   &           1   &         .86   &         .69   \\
$p$-value: SQ Shown+Fix $\times$ SQ Shown = SQ Opt+Fix $\times$ SQ Opt&         .32   &         .47   &         .74   &          .5   &         .95   \\
\end{tabular}

	 \flushleft \begin{footnotesize}
	 \begin{singlespace}

\textbf{Notes}:  This table contains regression specifications described in Section \ref{Specifications}. The exact wording of survey questions used as outcome variables is in Section \ref{SurveyQuestions}. ``Status Quo Shown (only if clicked)'' takes the value 1 if subjects have the option to view the status quo with an extra click, and choose to click, 0 otherwise.
	
	%\smallskip * significant at 10\%; ** significant at 5\%; *** significant at 1\%. 
	\end{singlespace}
	\end{footnotesize}
	\end{table}%

\begin{table}[h!]
\caption{\textbf{Status Quo Conditions and Demographics: Study 1 (``Scientific Veneer'')}}
	 \label{tab:SqDemogsStudy2}
	% \centering
	\scriptsize

\begin{tabular}{l*{14}{c}}
                    &\multicolumn{1}{c}{Positive}&\multicolumn{1}{c}{Rec (Scale)}&\multicolumn{1}{c}{Rec (Y/N)}&\multicolumn{1}{c}{Lawsuit}&\multicolumn{1}{c}{Damage Size}\\
\hline
Algorithm Fix       &         .34***&         .32***&    -2.9e-15   &        -.51***&        -.23***\\
                    &      (.055)   &      (.047)   &   (2.9e-09)   &      (.057)   &      (.042)   \\
Status Quo Shown    &         .35***&         .55***&         .51***&        -.21** &        -.16*  \\
                    &      (.086)   &      (.086)   &      (.081)   &      (.085)   &      (.086)   \\
Status Quo Shown (only if clicked)&         .28***&         .57***&         .57***&        -.27***&        -.21** \\
                    &      (.088)   &      (.087)   &       (.08)   &      (.084)   &      (.092)   \\
Fix $\times$ Status Quo Shown&        -.11   &        -.16** &     3.0e-15   &        .087   &         .03   \\
                    &      (.072)   &      (.064)   &   (2.3e-09)   &      (.079)   &      (.056)   \\
Fix $\times$  Status Quo Shown (only if clicked)&       -.043   &        -.14** &     5.5e-15***&        .089   &        .023   \\
                    &      (.074)   &      (.062)   &   (1.5e-15)   &      (.078)   &      (.062)   \\
Female              &        -.05   &       -.016   &       -.034   &       .0037   &          .1   \\
                    &      (.063)   &      (.067)   &      (.067)   &      (.066)   &      (.075)   \\
Black               &        -.55***&        -.42***&        -.28** &       -.032   &       -.048   \\
                    &       (.11)   &       (.12)   &       (.12)   &       (.17)   &       (.19)   \\
Asian               &        -.17   &        -.12   &         -.2   &       -.053   &         .21*  \\
                    &       (.11)   &       (.12)   &       (.15)   &       (.14)   &       (.12)   \\
Other Ethnicity     &        -.53***&        -.43** &        -.19   &          .2   &         .14   \\
                    &       (.15)   &       (.17)   &       (.15)   &       (.15)   &       (.18)   \\
Political Conservatism (Standardized)&         .16***&         .13***&        .091***&        -.15***&        -.14***\\
                    &      (.032)   &      (.033)   &      (.034)   &      (.033)   &       (.04)   \\
AI Decisions        &        .019   &      -.0022   &       -.058   &      -.0055   &       -.028   \\
                    &      (.064)   &      (.067)   &      (.069)   &      (.068)   &      (.077)   \\
Prepared by Educ.   &         .17***&         .08   &        .062   &         -.2***&        -.16** \\
                    &      (.064)   &      (.068)   &      (.073)   &      (.067)   &      (.077)   \\
Knows ML            &         .12*  &         .15** &         .13*  &       -.027   &       -.098   \\
                    &      (.065)   &      (.067)   &      (.068)   &      (.069)   &      (.078)   \\
\hline
Observations        &       1,984   &       1,984   &       1,984   &       1,984   &       1,984   \\
\(R^{2}\)           &         .11   &         .11   &        .086   &        .094   &        .063   \\
\hline $p$-value: Status Quo (SQ) Shown + Fix $\times$ SQ Shown=0&       .0038   &     8.7e-06   &     9.4e-10   &         .19   &         .16   \\
$p$-value: SQ Optional + Fix $\times$ SQ Opt=0&       .0032   &     5.5e-07   &     3.5e-12   &        .043   &        .058   \\
\hline $p$-value: SQ Shown = SQ Opt&         .41   &         .81   &          .4   &         .45   &         .55   \\
$p$-value: Fix $\times$ SQ Shown = Fix $\times$ SQ Opt&         .31   &         .76   &           1   &         .98   &         .91   \\
$p$-value: SQ Shown+Fix $\times$ SQ Shown = SQ Opt+Fix $\times$ SQ Opt&         .99   &         .63   &          .4   &         .48   &         .53   \\
\end{tabular}

	 \flushleft \begin{footnotesize}
	 \begin{singlespace}

\textbf{Notes}:  This table contains regression specifications described in Section \ref{Specifications}. The exact wording of survey questions used as outcome variables is in Section \ref{SurveyQuestions}. ``Status Quo Shown (only if clicked)'' takes the value 1 if subjects have the option to view the status quo with an extra click, and choose to click, 0 otherwise.
	
	%\smallskip * significant at 10\%; ** significant at 5\%; *** significant at 1\%. 
	\end{singlespace}
	\end{footnotesize}
	\end{table}%

\begin{table}[h!]
\caption{\textbf{ Effects of Activism: Study 1 (``Counterfactual'' and ``Fatalism'' Activism)}}
	 \label{tab:EffectsOfRhetoricStudy1}
	% \centering
	\scriptsize

	\emph{Panel A: Effects on Adoption Decisions} \\ 
\begin{tabular}{l*{14}{c}}
                    &\multicolumn{1}{c}{Positive}&\multicolumn{1}{c}{Rec (Scale)}&\multicolumn{1}{c}{Rec (Y/N)}&\multicolumn{1}{c}{Lawsuit}&\multicolumn{1}{c}{Damage Size}\\
\hline
Hiring              &       -.088   &       -.055   &     2.0e-17   &        -.04   &        -.28***\\
                    &      (.082)   &      (.085)   &         (.)   &      (.087)   &      (.079)   \\
Algorithm Fix       &         .37***&         .27***&        .019   &        -.61***&        -.28***\\
                    &       (.04)   &      (.034)   &      (.034)   &      (.045)   &      (.034)   \\
Counterfactual Op-Ed&         .24** &         .19*  &         .14** &        -.12   &        -.14   \\
                    &       (.11)   &       (.11)   &      (.054)   &       (.12)   &       (.11)   \\
AI Fatalism Op-Ed   &        -.16   &        -.28***&       -.079   &         .19*  &        .061   \\
                    &       (.11)   &       (.11)   &      (.063)   &       (.11)   &      (.094)   \\
Fix$\times$ Counterfactual&       -.042   &      -.0064   &        -.28** &         .16*  &        .041   \\
                    &      (.078)   &      (.061)   &       (.11)   &      (.086)   &      (.075)   \\
Fix$\times$ AI Fatalism&       -.044   &        .012   &         .16   &          .1   &        .012   \\
                    &      (.075)   &      (.062)   &       (.13)   &      (.077)   &      (.058)   \\
\hline
Fixed Effects       &     Subject   &     Subject   &     Subject   &     Subject   &     Subject   \\
Observations        &       1,992   &       1,992   &       1,992   &       1,992   &       1,992   \\
\(R^{2}\)           &         .68   &         .72   &         .76   &         .63   &         .73   \\
$p$-value: Counterfactual (CF)=Fatalism (FAT)&      .00061   &      .00003   &      .00079   &       .0057   &        .051   \\
\hline $p$-value: Fix$\times$CF=Fix$\times$FAT&         .98   &         .82   &      .00079   &         .59   &         .74   \\
$p$-value: CF+Fix$\times$CF=FAT+Fix$\times$FAT&      .00016   &     .000021   &      .00079   &        .031   &         .11   \\
\end{tabular}

	\ \\ \

	\emph{Panel B: Effect on the Use of Counterfactual Information}: \\
\begin{tabular}{l*{14}{c}}
                    &\multicolumn{1}{c}{Rec (Scale)}&\multicolumn{1}{c}{Rec (Scale)}&\multicolumn{1}{c}{Damage Size}&\multicolumn{1}{c}{Damage Size}\\
\hline
Status Quo Seen (Instrumented by Op-Ed)&         1.2***&         .69***&        -.84***&        -.35*  \\
                    &       (.28)   &        (.2)   &       (.32)   &       (.21)   \\
\hline
Instrument          &    CF Op-Ed   &   FAT Op-Ed   &    CF Op-Ed   &   FAT Op-Ed   \\
Observations        &       1,992   &       1,992   &       1,992   &       1,992   \\
1st Stage F-Stat    &          86   &         187   &          86   &         187   \\
\end{tabular}

	 \flushleft \begin{footnotesize}
	 \begin{singlespace}

\textbf{Notes}:  This table contains regression specifications described in Section \ref{Specifications}. The exact wording of survey questions used as outcome variables is in Section \ref{SurveyQuestions}. Insofar as the op-eds influence choices directly rather than through the examining of the status quo, these instruments may exhibit exclusion restriction limits. Our treatments to always/never/optionally show the status quo measure allow us to measure these direct effects. 
	
	%\smallskip * significant at 10\%; ** significant at 5\%; *** significant at 1\%. 
	\end{singlespace}
	\end{footnotesize}
	\end{table}%

\begin{table}[h!]
\caption{\textbf{Effects of Activism: Study 2 (``Scientific Veneer'')}}
	 \label{tab:EffectsOfRhetoricStudy2}
	% \centering
	\scriptsize

	\emph{Panel A: Effects on Adoption Decisions} \\ 
\begin{tabular}{l*{14}{c}}
                    &\multicolumn{1}{c}{Positive}&\multicolumn{1}{c}{Rec (Scale)}&\multicolumn{1}{c}{Rec (Y/N)}&\multicolumn{1}{c}{Lawsuit}&\multicolumn{1}{c}{Damage Size}\\
\hline
Hiring              &        -.13   &        -.23** &    -1.3e-17   &       .0022   &        -.17*  \\
                    &      (.096)   &      (.094)   &         (.)   &      (.091)   &      (.093)   \\
Algorithm Fix       &         .35***&         .27***&        .049   &        -.52***&        -.24***\\
                    &       (.04)   &      (.036)   &      (.044)   &      (.043)   &      (.036)   \\
Scientific Veneer   &        .036   &        .093   &        -.21***&        -.17   &       -.036   \\
                    &       (.16)   &       (.15)   &      (.079)   &       (.15)   &       (.13)   \\
Pro-AI Argument     &         1.1***&         1.1***&         .17** &        -.86***&        -.48***\\
                    &       (.13)   &       (.14)   &      (.071)   &       (.14)   &       (.13)   \\
Scientific Veneer $\times$ Pro AI&       -.019   &       -.072   &         .27** &         .24   &       -.037   \\
                    &       (.21)   &        (.2)   &       (.11)   &       (.21)   &       (.19)   \\
Fix $\times$ Scientific Veneer&        .078   &       -.046   &         .43***&       .0013   &      -.0084   \\
                    &        (.1)   &       (.08)   &       (.16)   &        (.1)   &       (.07)   \\
Fix $\times$ Pro-AI &        -.29***&        -.28***&        -.34** &         .24** &        .033   \\
                    &      (.073)   &      (.062)   &       (.14)   &       (.11)   &       (.07)   \\
Fix $\times$ Scientific Veneer $\times$ Pro-AI&       -.084   &         .23*  &        -.54** &        .049   &          .1   \\
                    &       (.14)   &       (.12)   &       (.23)   &       (.16)   &       (.11)   \\
\hline
Fixed Effects       &     Subject   &     Subject   &     Subject   &     Subject   &     Subject   \\
Observations        &       1,984   &       1,984   &       1,984   &       1,984   &       1,984   \\
\(R^{2}\)           &         .62   &         .67   &         .63   &         .62   &         .69   \\
\hline $p$-value: Veneer (Ven) +Ven$\times$ProAI = 0&          .9   &         .87   &         .49   &         .66   &         .58   \\
$p$-value: Ven$\times$Fix +Ven$\times$ProAI$\times$Fix = 0&         .96   &        .027   &         .49   &         .69   &         .29   \\
$p$-value: Ven+Ven$\times$Fix +Ven$\times$ProAI+Ven$\times$ProAI$\times$Fix = 0&         .93   &         .11   &         .49   &         .36   &         .89   \\
\end{tabular}

	\ \\ \

	\emph{Panel B: Effect on the Use of Counterfactual Information}: \\
\begin{tabular}{l*{14}{c}}
                    &\multicolumn{1}{c}{Views Status Quo}\\
\hline
Scientific Veneer   &        -.39   \\
                    &       (.26)   \\
Pro-AI Argument     &        -.45*  \\
                    &       (.26)   \\
Scientific Veneer $\times$ Pro AI&         .77*  \\
                    &       (.43)   \\
\hline
Observations        &         334   \\
\(R^{2}\)           &        .013   \\
\end{tabular}

%ScientificVeneer-Table.tex
%StatusQuoScientific.tex

	 \flushleft \begin{footnotesize}
	 \begin{singlespace}

\textbf{Notes}: This table contains regression specifications described in Section \ref{Specifications}. The exact wording of survey questions used as outcome variables is in Section \ref{SurveyQuestions}.
	
	%\smallskip * significant at 10\%; ** significant at 5\%; *** significant at 1\%. 
	\end{singlespace}
	\end{footnotesize}
	\end{table}%

\clearpage

\section{Regression Specifications}\label{Specifications}

\subsection{Study 1:``Counterfactual'' and ``Fatalism''}
	\begin{equation}
	\begin{split}
Y_{i,t,fix}=& \beta[1(t=1)+1(fix=1)+Hiring+FE_{i}+CF_{i,t}+\\
& FAT_{i,t}+CFxFix_{i,t,fix}+FATxFix_{i,t,fix}]+\epsilon
\end{split}
\end{equation}

Where: 
\begin{itemize}
\item $Y_{i,t,fix}$ refers to the response to question $Y$ individual $i$ on case $t$ regarding before/after the 6-month effort to $fix$. We run separate regressions for each of the five questions categories. 
\item $i$ indexes subjects (approximately 500 each for Study 1 and 2). 
\item $t\in\{1,2\}$ indexes the first or second order. $\beta_{1(t=1)}$ is a fixed effect for the first period. 
		\item $fix\in{0,1}$ differentiates answers to the questions before ($fix=0$) or after ($fix=1$) the six-months of dedicated effort. $\beta_{1(fix=1)}$ is a fixed effect for after the fix. 
		\item $Hiring$ is equal to 1 for the hiring business case. 
		\item $FE_{i}$ refers to the subject-level fixed effect. Note that this subsumes the effects of a) the order of the two cases, as well as b) whether the status quo details were shown, hidden or optional. 
		\item $CF$ refers to the ``counterfactual'' condition. This is 1 for the subjects reading the counterfactual in the second case, and 0 for everyone in the first case.  
		\item $FAT$ refers to the ``fatalism'' condition. This is 1 for the subjects reading the fatalism in the second case, and 0 for everyone in the first case. 
		\item Note that no op-ed is the excluded condition. 
\item $CFxFix$ is equal to $CF\times 1(fix=1)$
	\item $FATxFix$ is equal to $FAT\times 1(fix=1)$
	\item $\epsilon$: Standard errors are clustered by subject. 
\end{itemize}

\subsection{Study 2: ``Scientific Veneer''}
	\begin{equation}
	\begin{split}
Y_{i,t,fix}= & \beta[1(t=1)+1(fix=1)+Hiring+FE_{i}+ProAi_{i,t}+SciVen_{i,t}+ProAixSciVen_{it}+\\ 
& ProAixFix_{i,t}+SciVenxFix_{i,t}+ProAixSciVenxFix_{it}]+\epsilon
\end{split}
\end{equation}

Where: 
\begin{itemize}
\item $Y_{i,t,fix}$ refers to the response to question $Y$ individual $i$ on case $t$ regarding before/after the 6-month effort to $fix$. We run separate regressions for each of the five questions categories. 
\item $i$ indexes subjects (approximately 500 each for Study 1 and 2). 
\item $t\in\{1,2\}$ indexes the first or second order. $\beta_{1(t=1)}$ is a fixed effect for the first period. 
		\item $fix\in{0,1}$ differentiates answers to the questions before ($fix=0$) or after ($fix=1$) the six-months of dedicated effort. $\beta_{1(fix=1)}$ is a fixed effect for after the fix. 
		\item $Hiring$ is equal to 1 for the hiring business case. 
		\item $FE_{i}$ refers to the subject-level fixed effect. Note that this subsumes the effects of a) the order of the two cases, as well as b) whether the status quo details were shown, hidden or optional. 
		\item $ProAi$ refers to the condition in which a positive argument about AI ethics is made. This is 1 for the subjects reading the positive opinion in the second business case, and 0 for everyone in the first case.  
		\item $SciVen$ refers to whether the expert applied scientific veneer. This is 1 for the subjects seeing scientific veneer in the second case, and 0 for everyone in the first case. 
		\item $ProAixSciVen$ refers to subjects who see pro-AI arguments with scientific veneer. This is 1 for the subjects reading these arguments in the second case, and 0 for everyone in the first case. 
\item $ProAixFix$ is equal to $ProAi\times 1(fix=1)$
	\item $SciVen$ is equal to $SciVen\times 1(fix=1)$
	\item $ProAixSciVenxFix$ is equal to $ProAi\times SciVen\times 1(fix=1)$
	\item Note that anti-AI, excluded is the omitted condition. 
	\item $\epsilon$: Standard errors are clustered by subject. 
\end{itemize}

\section{Experimental Materials: Case Studies}\label{CaseStudyTexts}

\subsection{Text of Hiring-Related Business Case}

The focus in this business case is on one of the world's top technology companies, headquartered in California (think of Facebook, Alphabet, Apple...). The case refers to it as "Toptech Co.".

Please imagine you are an engineering executive for Toptech Co. The organization you lead at Toptech implements software behind Toptech’s consumer-facing products, digital infrastructure and advertising revenue.

As part of your job, you need to recruit and hire workers. Hiring qualified technology workers is difficult and high-stakes. Hiring quality people can make an enormous difference in product quality, market leadership and profitability. Many technology companies have reported shortages of qualified workers.

Toptech gets high volumes of applications from aspiring software engineers from around the world. Because of this high volume, it is cost-prohibitive to hire recruiters to read each job application with great care. As a result, a team at Toptech was assigned to work with your team to develop software to read the text of job applications and score them.

The technology looks at Toptech’s historical hiring and job performance records. This data contains information about what job candidates and workers perform well on the job, who doesn’t, and what each candidates’ personal characteristics, experiences and qualifications were on the job application

Using this data, a team of statistical and machine learning experts has built a sophisticated model of what candidates are is likely to succeed using this data. The model is truly sophisticated; it allows for many different “paths” to being labeled a likely success. In data from the past five years, the model is very accurate at predicting which candidates will succeed or fail, even as circumstances have changed in the tech industry and at Toptech.

Advocates of the program claim that it improves the objectivity and consistency of evaluation. However, critics are concerned that the algorithm may propagate biases in the historical data. For example, critics inside the company point to the fact that the software’s recommendations for who to hire in technical roles are approximately 71\% men, and 29\% women.

Your questions today are about whether your organization should adopt this technology.

\subsection{Text of Lending-Related Business Case}

For this scenario, imagine you are a finance executive at one of the largest banks in the country (think of Bank of America, Wells Fargo, Citigroup...). The case will refer to it as "Financial Co." The organization you lead at Financial oversees consumer lending products and services. 

As part of your job, you need to oversee a process for making decisions about small business lending. This is a critical part of Financial's strategy. 

Financial gets high volumes of applications from small businesses. Because of this high volume, it is costly to hire loan officers to read each loan application with great care. As a result, a team at Financial was assigned to work with your team to develop software to read the content of loan applications and score them.

The technology looks at Financial's historical data about loan performance. This contains data about what borrowers historically paid back their loans, who didn't and what all applicants' loan characteristics are. 

Using this data, a team of statistical and machine learning experts has built a sophisticated model of what borrowers are likely to pay back using this data. The model is truly sophisticated; it allows for many different “paths” to being labeled a likely success. In data from the past five years, the model is very accurate at predicting which borrowers will pay back or not, even as circumstances have changed in the economy. 

Advocates of the program claim that it improves the objectivity and consistency of loan evaluation. However, critics are concerned that the algorithm may propagate biases in the historical data. For example, critics inside the company point to the fact that the software’s recommendations charge higher and more expensive interest rates -- about 3.2 higher basis points -- to Latinx and African-American borrowers, compared to white borrowers. 

Your questions today are about whether your organization should adopt this technology.

\section{Survey Questions}\label{SurveyQuestions}

\begin{itemize}
\item How negative or positive is the impact of this technology? (very negative - very positive)
\item What is the probability that a fairness issue is alleged and becomes a problem (through lawsuits or PR)? (very low - very high)
\item If a problem is alleged, how damaging would that be? (not at all damaging - very damaging)
\item To what extent do you recommend using this algorithm rather than the status quo processes? (Do not recommend - Strongly recommend)
\item If you were forced to make a decision on the future use of the algorithm now, what would you do? (Use the status quo methods (don't switch to the algorithm) OR Switch to adopt this algorithm).
\end{itemize}

[pagebreak]

Now suppose that the team agrees to spend the next six months making an effort to address fairness issues. Please re-evaluate the proposal based on how you expect the algorithm to perform after six additional months of your team attempting to address fairness issues.

\section{Experimental Materials: ``Scientific Veneer'' Interventions}\label{MaterialsVeneer}

\subsection{Expert Introductions}

For our veneer treatment, the following introduction was provided to the expert: \ \\ \

\emph{To help your team make a decision, your company has hired an expert to offer an assessment of the technology.}  

\emph{The expert, Taylor, has a Ph.D. in Physics from UC Berkeley. Taylor has been employed as a third-party evaluator of digital technology for the past two years, working on questions like those you're facing.}

\emph{On the next page you will read Taylor's summary and analysis of the technology and decision you are facing.} \ \\ \ 

For our non-veneer treatment, the following introduction was provided to the expert: \ \\ \

\emph{To help your team make a decision, your company has hired an expert to offer an assessment of the technology.} 

\emph{The expert, Taylor, has a PhD. in Sociology from UC Berkeley. Taylor has been employed as a third-party evaluator of digital technology for the past two years, working on questions like those you're facing.}

\emph{On the next page you will read Taylor's summary and analysis of the technology and decision you are facing.}

\subsection{Pro-AI Argument Without Scientific Veneer}

\begin{center}
    \includegraphics[width=10cm]{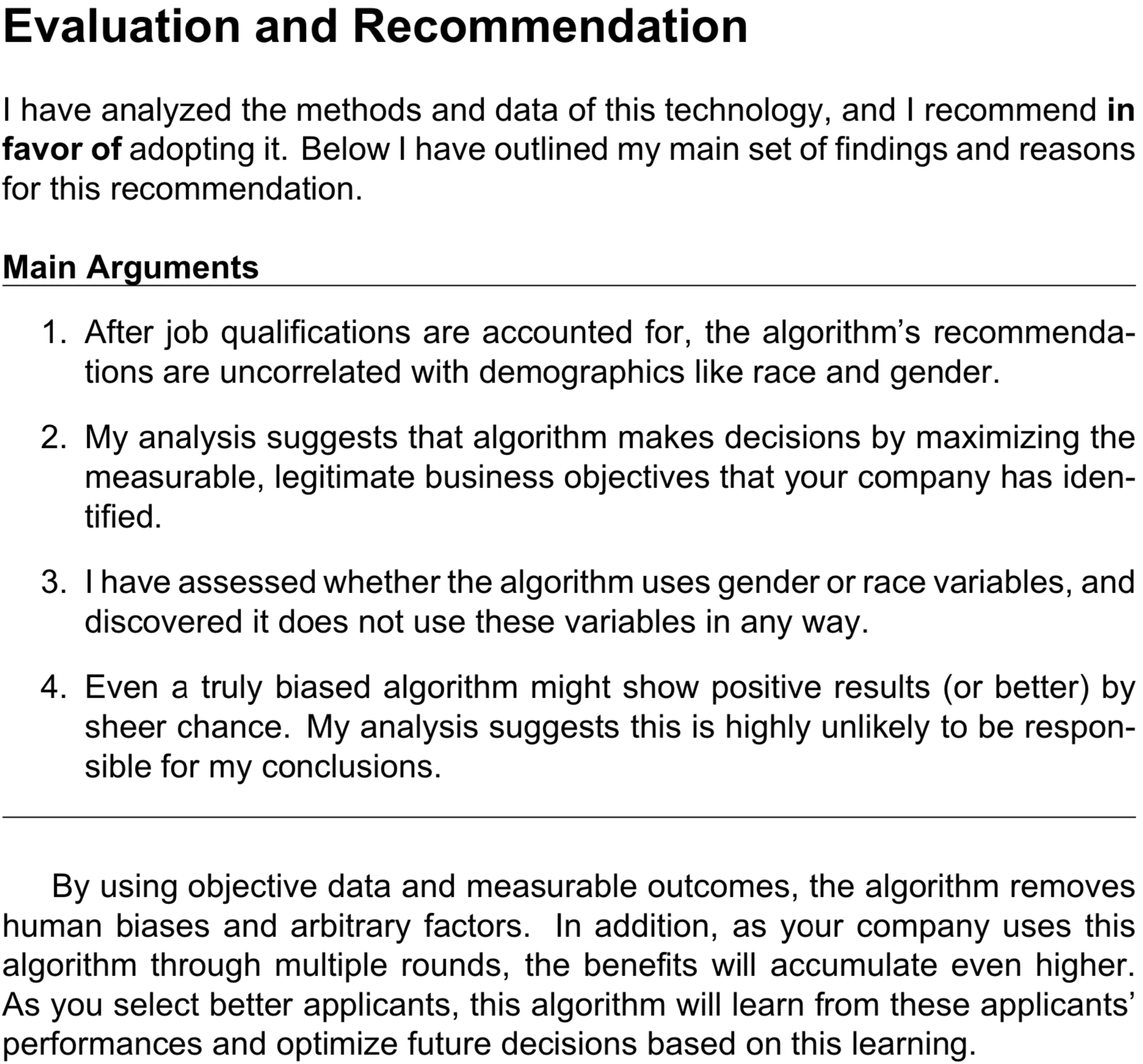}
\end{center}

\clearpage
\subsection{Pro-AI Argument With Scientific Veneer}

\begin{center}
    \includegraphics[width=10cm]{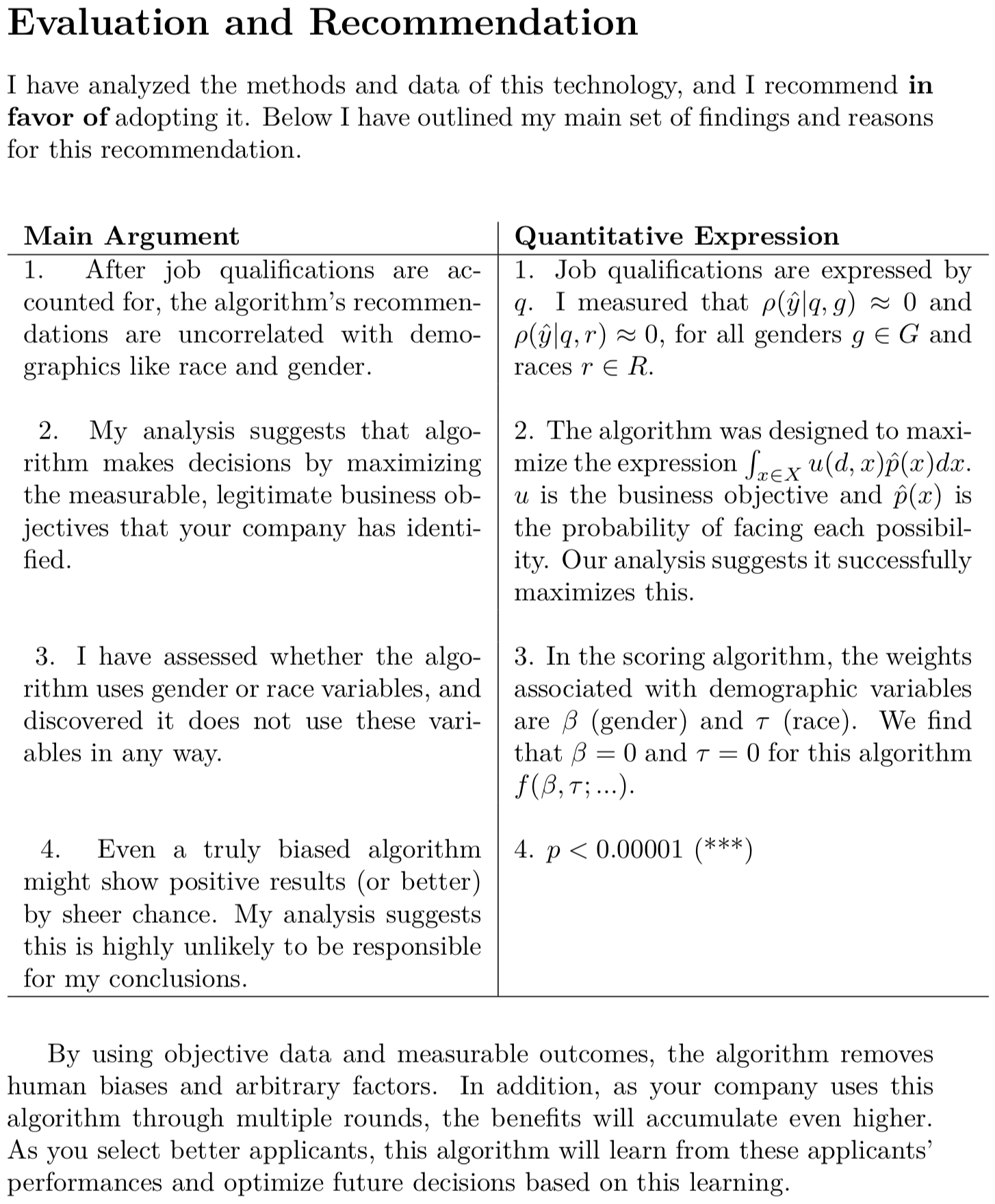}
\end{center}

\clearpage
\subsection{Anti-AI Argument Without Scientific Veneer}

\begin{center}
    \includegraphics[width=10cm]{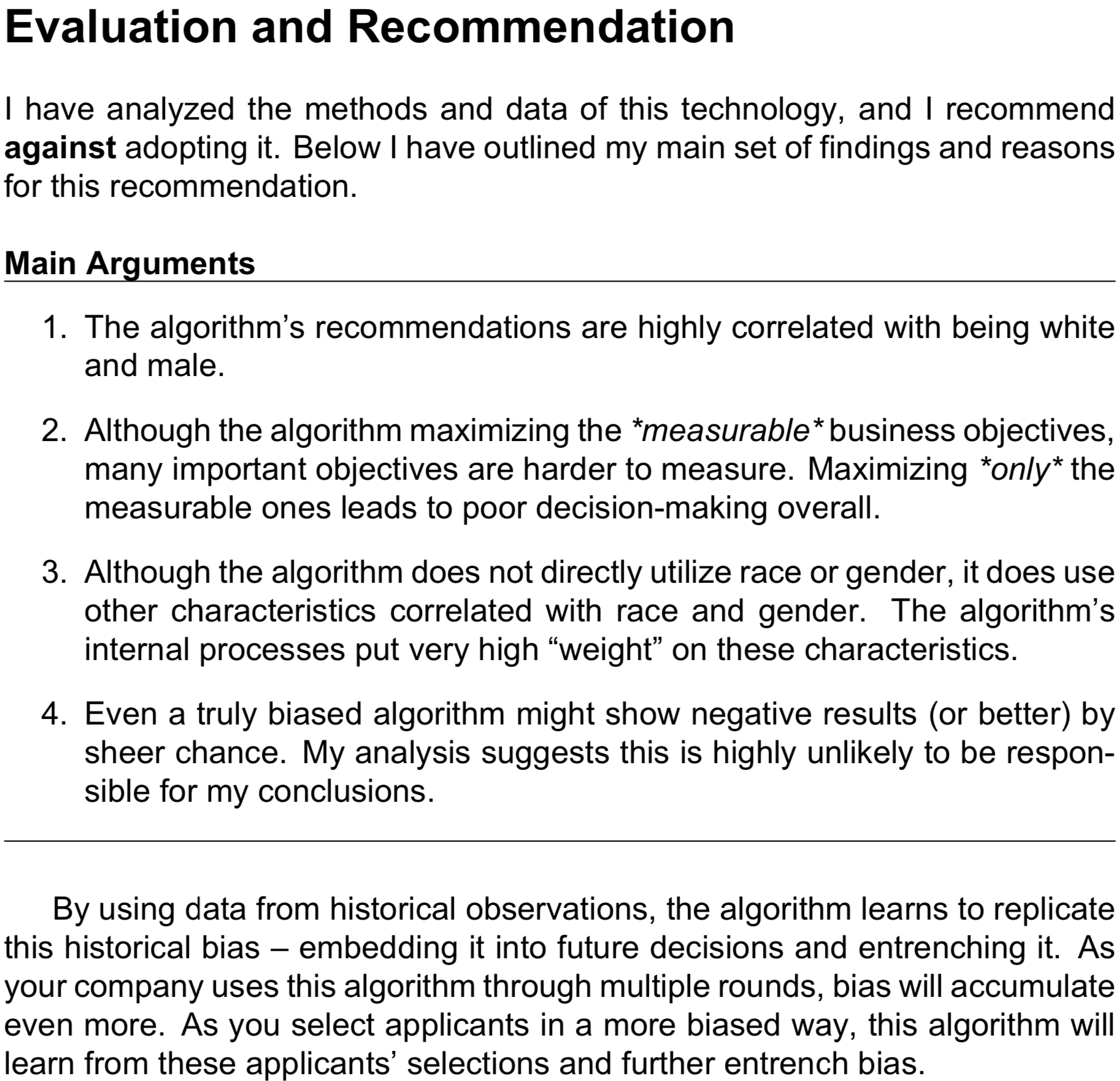}
\end{center}

\clearpage
\subsection{Anti-AI Argument With Scientific Veneer}

\begin{center}
    \includegraphics[width=10cm]{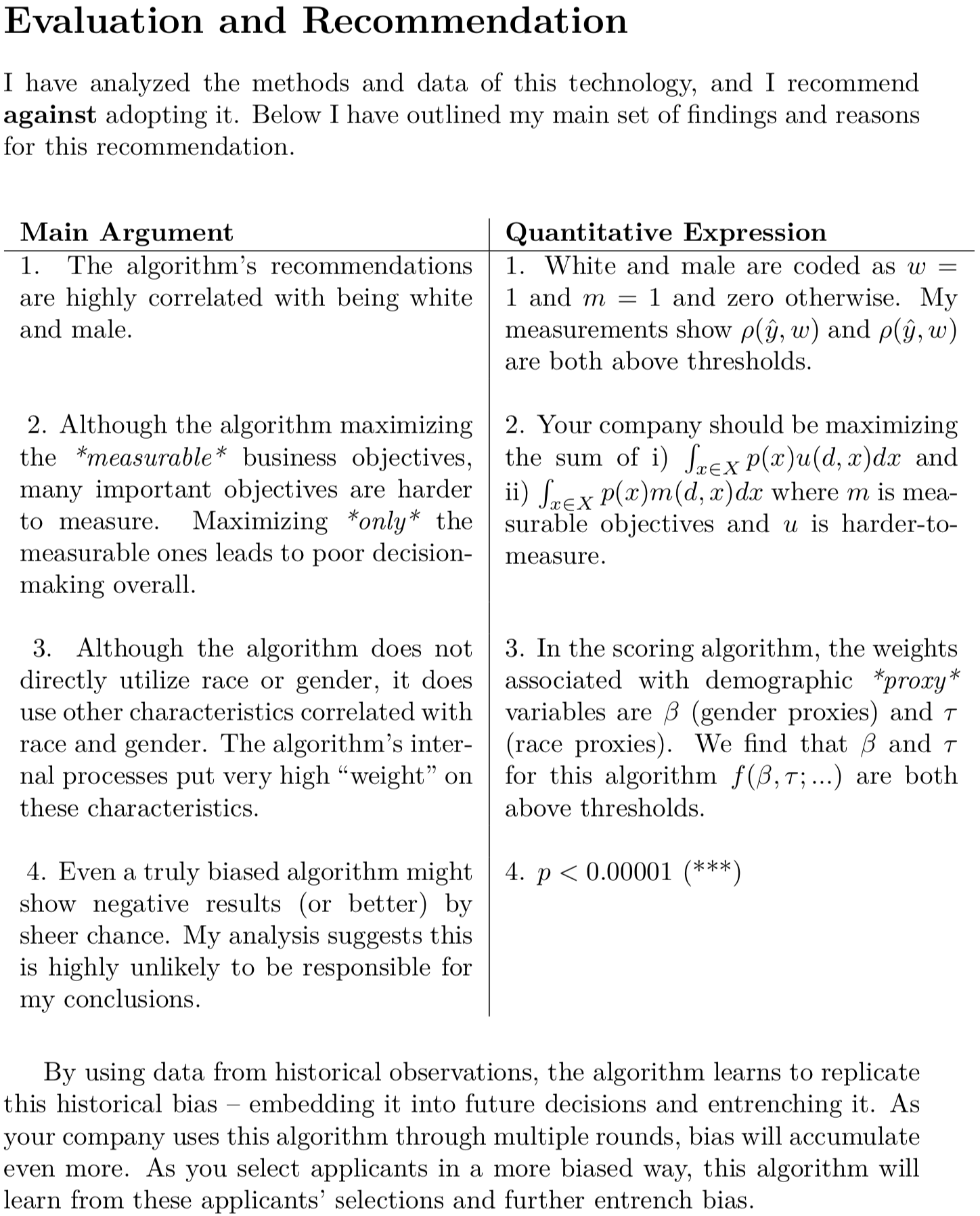}
\end{center}

\putbib
\end{bibunit}

\end{document}